\documentclass[12pt,preprint2]{aastex}

\usepackage{psfig}

\slugcomment{submitted to ApJ}

\shorttitle{X-ray and Optical Sources in NGC 1073}
\shortauthors{Kaaret}

\begin{document}

\title{Optical Sources Near the Bright X-Ray Source in NGC 1073}

\author{Philip Kaaret} 
\affil{Department of Physics and Astronomy, University of Iowa,  Van
Allen Hall, Iowa City, IA 52242, USA}


\begin{abstract}

New HST observations show that the bright X-ray source in the face-on
spiral galaxy NGC 1073 is located near a ring of recent star formation
with an age of 8--16 Myr.  This strengthens the association of X-ray
sources in spiral galaxies emitting near or above the Eddington limit
for a $20 M_{\odot}$ black hole with recent star formation events.  Two
candidate optical counterparts of the X-ray source are found.  The
X-ray to optical flux ratios of both are consistent with those of
low-mass X-ray binaries and higher than most high-mass X-ray binaries,
suggesting that reprocessing of X-rays contributes to the optical
light.  The optical magnitude and color of one candidate is consistent
with that predicted for an X-ray binary with an initial donor mass of
$6-8 \, M_{\odot}$.  However, the same X-ray binary evolution model
underestimates the X-ray luminosity.  An X-ray source list for the
field is presented which includes detections of the nucleus of NGC
1073, three quasars, and an M3e brown dwarf star with high proper
motion.

\end{abstract}


\section{Introduction}

The discovery of bright X-ray sources in external galaxies has
generated significant interest due to the possibility that they may be
`intermediate-mass' black holes suggested by the masses required to not
violate the Eddington limit for the inferred high luminosities
\citep{colbert99,makishima00,kaaret01}.  However, there are alternative
interpretations of these so-called `ultraluminous X-ray sources' (ULXs)
in which the X-rays are beamed or the Eddington limit is violated and
the existence of intermediate-mass black holes are not required.  The
luminosity function of X-ray point sources in external galaxies does
not appear to have a break around the Eddington limit for `normal'
stellar mass black holes, i.e. with $M < 20 M_{\odot}$, suggesting that
objects above and below this limit and members of the same population
\citep{kilgard02,grimm03}.  It is useful to study objects both below
and above this limit to understand the properties of the full
population.

The identification of counterparts of these X-ray sources at other
wavelengths is important to understand the physical nature of these
objects \citep{liu02,pakull02,kaaret03,zampieri04,liu04,kaaret04b}. 
Classification of the spectral types of the companion stars should
directly constrain the evolutionary history of the binary systems. 
Spectroscopy of companion stars might permit measurement of radial
velocity curves providing direct constraints on the compact object
mass.  Characterization of the environments in which the X-ray sources
are found should provide clues to their formation
\citep{kaaret04a,soria04}.

Here, we report on Hubble Space Telescope and Chandra X-Ray Observatory
observations of the face-on spiral galaxy NGC 1073 (= UGC 2210).  This
galaxy contains an ``Intermediate X-ray Object'', IXO 5, reported in
the catalog of \citet{colbert02}.  This object has an X-ray luminosity
of $\sim 2 \times 10^{39} \rm \, erg \, s^{-1}$ which is below the
Eddington luminosity for a $20 M_{\odot}$ black hole.  However, it is
significantly brighter than any persistent black hole X-ray binary in
the Milky Way and lies at the transition between standard black hole
X-ray binaries and ultraluminous X-ray sources.

NGC 1073 is a member of a tight group of galaxies containing the bright
Seyfert galaxy NGC 1068 and some additional fainter companions.  We
adopt a distance to NGC 1073 of 16.4~Mpc based on a radial velocity
corrected for infall of the local group toward Virgo of 1147 km/s as
reported in the LEDA catalog and a Hubble constant of 70~km/s/Mpc.  NGC
1073 is notable because several quasars lie near the light of sight
\citep{arp79}.  The presence of several objects in the field with both
X-ray and optical emission permits us to obtain accurate relative
astrometry of the X-ray and optical images.  There are only two
potential optical counterparts to the brightest X-ray source in the
galaxy.  We describe the observations and analysis in \S~2, and discuss
the results in \S~3.

\section{Observations and Analysis}

\subsection{HST observations}

Observations of NGC 1073 were made using the Advanced Camera for
Surveys (ACS) on the Hubble Space Telescope (HST) under GO program
10001 (PI Kaaret).  Images were obtained in the broad band filters
F435W (Johnson B) and F606W (Broad V) using the Wide-Field Camera
(WFC).  All the observations were made on 18 Nov 2003.  The pointing
was offset from either the X-ray source position or the galaxy nucleus
in order to include two quasars known to emit both optical light and
X-rays in the ACS field of view to allow us to align the X-ray and
optical images.  Each observation consisted of a two point line dither
pattern with a pair of cosmic-ray split images obtained at each point
in the pattern.  The total exposure was 2160~s for the F435W image and
2240~s for the F606W image.

We used the images delivered by the standard ACS pipeline processing
(OPUS 15.a and CALACS code version 4.4.1) which removes cosmic-rays,
corrects for optical distortion, and dither combines the images. We
aligned the F435W image to stars in the USNO B1.0 catalog
\citep{usnob1} using the {\it imwcs} tool from the Smithsonian
Astrophysical Observatory Telescope Data Center.  Many of the stars in
the USNO B1.0 catalog appear as multiple stars or clusters in the HST
image.  We selected 10 stars which appear as single stars in the HST
image and have counterparts in the USNO B1.0 catalog.  The standard
deviation of the offsets of the stellar positions is $0.37\arcsec$
which is comparable to the expected accuracy of the USNO B1.0
positions.  With ten reference stars, the absolute astrometry of the
corrected HST image should be better than $0.2\arcsec$.  The F606W
image was aligned to the aspect-corrected F435W image  using the
{\it IRAF} tools {\it geomap} and {\it geotran} \citep{tody93}.

\begin{figure*}[th] \centerline{\epsscale{2.0}\plotone{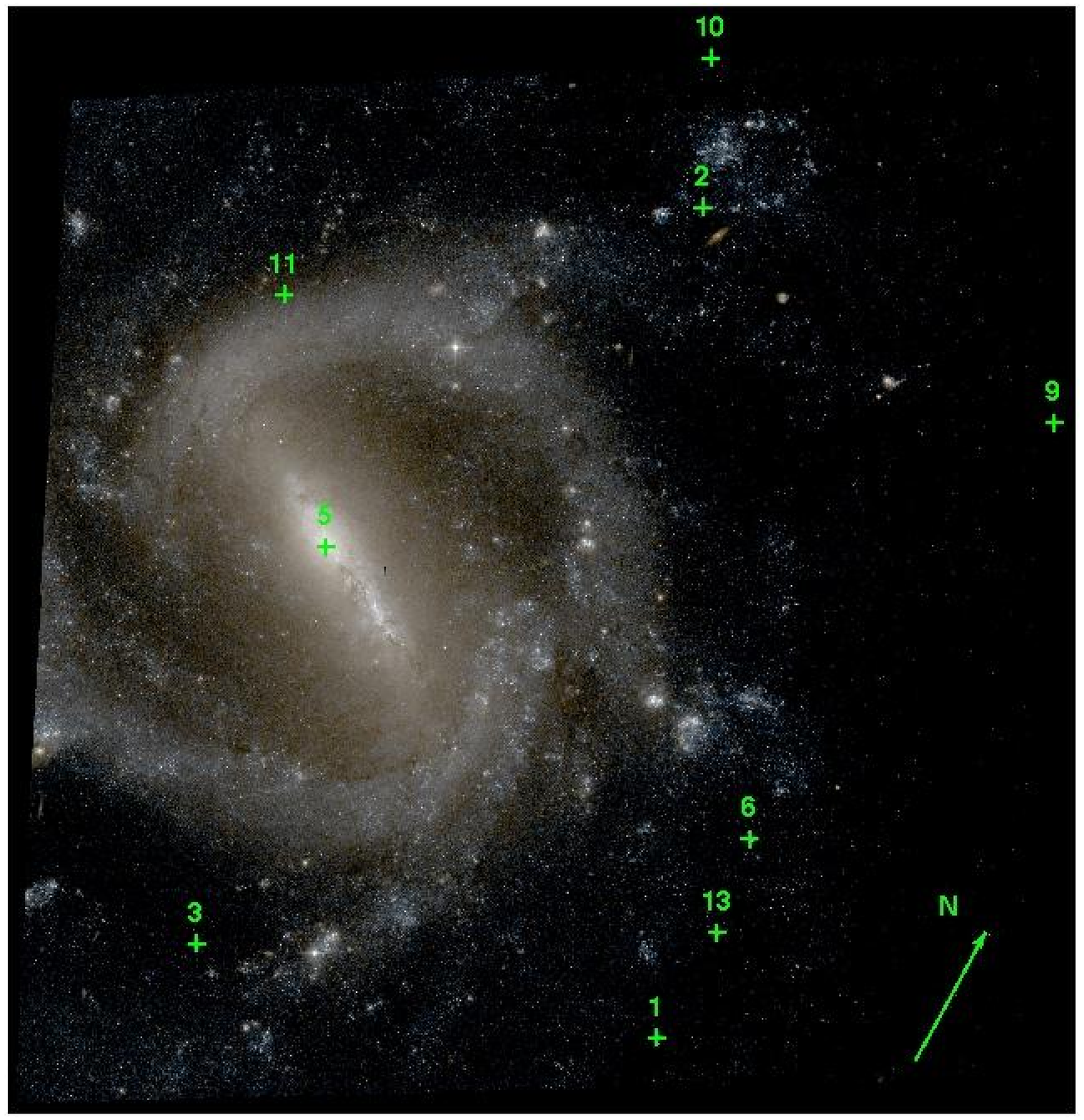}}
\caption{NGC 1073 imaged with the HST/ASC.  Counts from the F435W
filter image appear as blue; counts from the F606W filter image appear
as red.  A green color channel was interpolated between the F435W and
F606W images to approximate true colors.  The green crosses indicate
the positions of Chandra sources as numbered in Table~\ref{xsources}. 
Note that some of the Chandra sources lie outside the HST/ASC field of
view and are not present on the image.  The green arrow points North
and has a length of $10\arcsec$.\label{bigimage}} \end{figure*}

\begin{deluxetable}{rrrcccrrl}
\tabletypesize{\scriptsize}
\tablecaption{X-ray sources near NGC 1073
  \label{xsources}}
\tablewidth{0pt}
\tablehead{ &
  \colhead{S/N} & 
  \colhead{Counts} & 
  \colhead{RA} & \colhead{DEC} & \colhead{Error} &
  \colhead{Flux}  &
  \colhead{Hardness}  &
  \colhead{Comment} }
\startdata
 1 & 65.0 & 162 & 02 43 33.401 & +01 21 37.23 & 0.1 & $26.0 \pm 2.0$ & 0.45 & QSO B0240+011 \\
 2 & 22.5 &  44 & 02 43 38.158 & +01 24 11.28 & 0.1 &  $6.8 \pm 1.0$ & 0.73 & IXO 5 \\
 3 & 19.1 &  50 & 02 43 39.533 & +01 21 09.91 & 0.1 & $10.6 \pm 1.5$ & 0.74 & QSO B0241+011 \\
 4 & 14.6 &  34 & 02 43 37.852 & +01 29 05.95 & 0.2 &  $5.6 \pm 1.0$ & 0.54 & \\
 5 & 14.3 &  28 & 02 43 40.515 & +01 22 33.99 & 0.2 &  $4.3 \pm 0.8$ & 0.86 & Nucleus \\
 6 &  8.7 &  26 & 02 43 33.553 & +01 22 21.90 & 0.3 &  $4.0 \pm 0.8$ & 0.69 & J024333.6+012222 \\
 7 &  6.6 &  24 & 02 43 22.088 & +01 28 52.70 & 0.1 &  $4.4 \pm 0.9$ & 0.36 & LP 590-443 \\
 8 &  5.7 &  10 & 02 43 38.276 & +01 25 36.01 & 0.2 &  $1.7 \pm 0.5$ & 0.54 & \\
 9 &  4.7 &   8 & 02 43 32.560 & +01 24 06.19 & 0.2 &  $1.4 \pm 0.5$ & 0.89 & \\
10 &  4.7 &   9 & 02 43 39.010 & +01 24 39.07 & 0.2 &  $1.4 \pm 0.5$ & 1.00 & \\
11 &  4.6 &   9 & 02 43 42.627 & +01 23 15.51 & 0.1 &  $1.4 \pm 0.5$ & 0.33 & Star \\
12 &  4.6 &  19 & 02 43 57.150 & +01 18 33.95 & 0.4 &  $4.8 \pm 1.1$ & 0.85 & \\
13 &  3.5 &   7 & 02 43 33.347 & +01 22 01.89 & 0.2 &  $1.2 \pm 0.4$ & 0.87 & Star \\
\enddata

\vspace{-12pt}\tablecomments{Table~\ref{xsources} includes for each
source: the source number; S/N -- the significance of the source
detection as calculated by {\it wavdetect}; Counts - the net counts in
the 0.3--7~keV band; RA and DEC -- the position of the source in J2000
coordinates; Error - the statistical error on the source position in
arcseconds, note that this does not include errors in the overall
astrometry; Flux -- the source flux in units of $10^{-14} \, \rm erg \,
cm^{-2} \, s^{-1}$ in the 0.3--7~keV band calculated assuming a power
law spectrum with photon index of 1.5 and corrected for the Galactic
absorption column density of $4.2 \times 10^{20} \rm \, cm^{-2}$;
Hardness - the hardness ratio defined as the ratio of counts in the
1--7~keV band to counts in the 0.3--7~keV band.  The useful exposure
was 5.7~ks.}   \end{deluxetable}

\subsection{Chandra observations}

Observations of NGC 1073 were made using the Advanced CCD Imaging
Spectrometer spectroscopy array (ACIS-S; Bautz et al.\ 1998) onboard
the Chandra X-Ray Observatory.   The ACIS-S was used in imaging mode
and the source position for IXO 5 as reported by \citet{colbert02} was
placed at the aimpoint.  The {\it Chandra\/} observation (ObsID 4686;
PI Kaaret) began on 9 Feb 2004 08:16:29 UT and had a useful exposure of
5.7~ks.  The Chandra data were subjected to standard data processing
and event screening (ASCDS version 7.1.1 using CALDB version 2.25). 
The total rate on the S3 chip was below 1.1~c/s for the entire
observation indicating that there were no strong background flares. 

We constructed an image using all valid events on the S2 and S3 chips
and used the {\it wavdetect} tool which is part of the {\it CIAO}
version 3.1 data analysis package to search for X-ray sources.  The
list of detected sources with detection significance of $3.5\sigma$ or
higher is given in Table~\ref{xsources}.  The Chandra astrometry was
aligned to the HST astrometry as described below.  We computed an
exposure map for an assumed source spectrum of a powerlaw with photon
index of 1.5 absorbed by a column density of $4.2 \times 10^{20} \rm \,
cm^{-2}$ which is the total Galactic {\sc H i} column density along the
line of sight \citep{dickey90}.  In order to give some indication of
the spectral shape of each source, we calculate the the ratio of the
Chandra counts in the 1--7~keV band to counts in the 0.3--7~keV band.
We searched for X-ray variability by  comparing the photon arrival
times for each source to the distribution expected for a constant
source with the same average flux using a Kolmogorov-Smirnoff (KS)
test.  No background subtraction was performed.  Marginal evidence for
variability was found for two sources: IXO 5 (source \#2)  which is
variable at a confidence level of 94\% and the quasar J024333.6+012222
(source \#6) which is variable at a confidence level of 95\%.

Three quasars are detected with Chandra.  These are QSO B0240+011 and
QSO B0241+011 which were previously detected with ROSAT
\citep{colbert02} and J024333.6+012222 \citep{vv96}.  The nucleus of
NGC 1073 is detected at a luminosity (assuming isotropic emission) of
$1.4 \times 10^{39} \rm \, erg \, s^{-1}$ in the 0.3--7~keV band. 
There is an X-ray source (\#7) with a relatively soft spectrum within
$2\arcsec$ of the position of LP 590-443 which is a M3e brown dwarf
star with H$\alpha$ emission and high proper motion \citep{cook00}.

IXO 5 is clearly detected with a position offset by $2.9\arcsec$ from
that reported by \citet{colbert02}.  This is well within the ROSAT
position uncertainty.  We fit the X-ray spectrum of the source using
the {\it XSPEC} spectral fitting tool which is part of the {\it
LHEASOFT} X-ray data analysis package and response matrices calculated
using {\it CIAO}.  Not many photons were detected from IXO 5, so
constraints on the spectral parameters are poor.  An adequate fit was
obtained with an absorbed powerlaw spectrum.  The best fit parameters
were photon index of $1.3^{+0.8}_{-0.4}$ and an equivalent hydrogen
absorption column density of $N_{H} = 6^{+30}_{-2}\times 10^{20} \rm \,
cm^{-2}$.  The lower bound on $N_{H}$ was fixed to the Galactic {\sc H
i} column density along the line of sight.  The source flux from the
spectral fit is consistent with that listed in Table~\ref{xsources}.

\subsection{Optical counterparts to X-ray sources}

We overlaid the Chandra X-ray source positions on the aspect corrected
F435W image.  We found 6 probable matches between X-ray sources and
bright optical sources within the estimated uncertainty of 0.6$''$ at
90\% confidence  of the {\it Chandra} aspect solution (see the Chandra
aspect pages at http://cxc.harvard.edu/cal/ASPECT/celmon).  The Chandra
sources with optical counterparts are  sources \#1, \#3, \#5, \#6,
\#11, and \#13.  The first two sources are the X-ray emitting quasars
discussed earlier. Source \#3 lies on the ACIS-S2 chip rather than the
S3 chip where the bright X-ray source is located and we chose not to
use it for the astrometric solution.  Source \#6 is the nucleus of the
galaxy and is unsuitable for astrometry.  We used the remaining four
sources (\#1, \#5, \#11, and \#13) to align the Chandra astrometry to
that of the HST image (which had previously been aligned with the USNO
B1.0 catalog).  We estimate that the relative alignment is accurate to
better than $0.3\arcsec$.

\begin{figure}[t] \centerline{\epsscale{1.0}\plotone{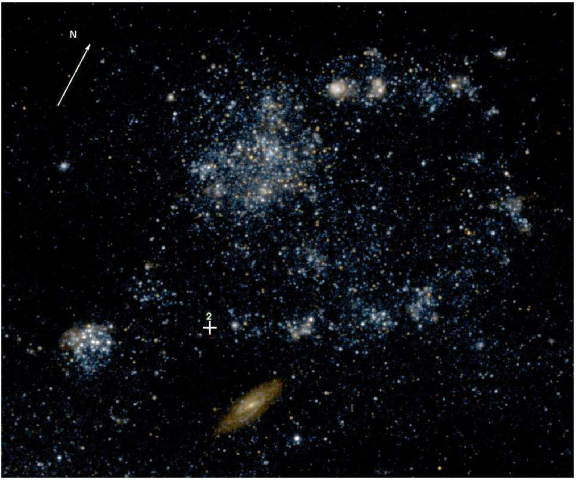}}
\caption{Image of the region near IXO 5.    The colors are as in
Fig.~\ref{bigimage}.  The white cross labeled ``2" shows the position
of IXO 5.  The white arrow points North and has a length of
$5\arcsec$.\label{sb_image}}   \end{figure}

\begin{figure}[t] \centerline{\epsscale{1.0}\plotone{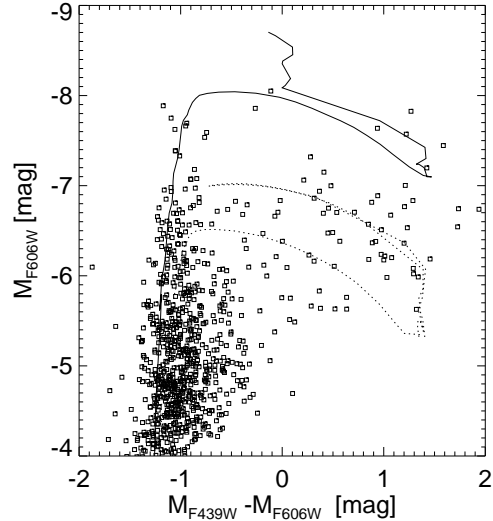}}
\caption{Color-magnitude diagram in $M_{\rm F606W}$ versus $M_{\rm
F439W} - M_{\rm F606W}$ for the field of Fig.~\ref{sb_image}.  The
magnitudes are ST magnitudes calculated for the bandpasses of the
HST/WPFC2 F439W and F606W filters.  The solid curve is an isochrone for
an 8~Myr old stellar population from \citet{bertelli94} for a
metallicity of $z = 0.019$.  The dotted curve is an  isochrone for a
16~Myr old stellar population. \label{sb_hr}}   \end{figure}

An optical image of the region near IXO 5 is shown in 
Fig.~\ref{sb_image}.  IXO 5 appears to lie near a star-forming ring. 
We performed photometry of the entire field shown in
Fig.~\ref{sb_image} using DAOPHOT Classic \citep{stetson87} as
translated into the Interactive Data Language (IDL) by W.B. Landsman. 
We applied an intensity threshold somewhat above the limiting magnitude
of the images in order to select only sources for which good photometry
could be obtained.  We selected only sources for which the absolute
value of the DAOPHOT peak parameter was less than 0.75 to remove
extended sources and image defects and also required that the fit of
the model point spread function to source intensity profile have
$\chi^2 < 3$.  These cuts remove a number of sources, including some
with $M_V \sim -10$ which are likely H{\sc ii} regions.  However, some
unresolved or marginally resolved H{\sc ii} regions or star clusters
may remain.

To enable direct comparison with theoretical stellar isochrones, see
below, we calculated ST magnitudes in the F439W and F606W filter
bandpasses of the HST Wide-Field Planetary Camera 2 (WFPC2).  The
reddening along the line of sight to NGC 1073 calculated from dust maps
derived from COBE data \citep{schlegel98} gives an extinction $\rm
E(B-V) = 0.039$.  Using this extinction, a Galactic reddening law, and
the {\it synphot} package which is part of the {\it STSDAS} HST data
analysis software under {\it IRAF}, we calculated reddening corrected
magnitudes from the count rates measured in the F435W and F606W filters
using the spectrum of a A0V star taken from the Bruzual stellar
spectrum atlas available in {\it synphot}.  Because the WFPC2 F439W
filter is well matched to the ACS/WFC F435W filter and the WFPC2 F606W
filter is well matched to the ACS/WFC F606W filter, translation of the 
ACS/WFC count rates into ST magnitudes for the WFPC2 filter bandpasses
is not strongly dependent on color.  From the results presented by
\citet{sirianni05}, we estimate that the maximum error on the ST
magnitudes induced by the transformation is 0.14 for $m_{\rm F439W}$
and 0.11 for $m_{\rm F606W}$.

A color-magnitude diagram of $M_{\rm F606W}$ versus $M_{\rm F439W} -
M_{\rm F606W}$ for the field of Fig.~\ref{sb_image} is shown in
Fig.~\ref{sb_hr}.  The magnitudes are ST magnitudes calculated for the
bandpasses of the HST/WPFC2 F439W and F606W filters.  The colors are
typically accurate to within $\pm 0.4$.  There may be additional
reddening within NGC 1073 or internal to the source which would affect
the magnitudes and colors.  Shown on the figure are isochrones for
stellar populations for a metallicity of $z = 0.019$ with ages of 8 and
16 Myr \citep{bertelli94,girardi02}.

\begin{figure}[t] \centerline{\epsscale{1.0}\plotone{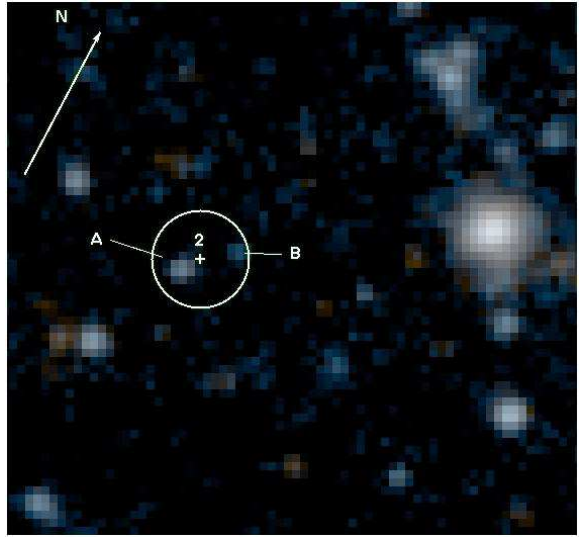}}
\caption{Image of the region near IXO 5.  The colors are as in
Fig.~\ref{bigimage}.  The white cross labeled ``2" shows the position
of IXO 5.  The white circle has a radius of $0.3\arcsec$ and represents
the relative Chandra/HST error circle.  The white arrow points North
and has a length of $2\arcsec$.\label{ixoclose}}   \end{figure}

Fig.~\ref{ixoclose} shows the candidate counterparts found to IXO 5 in
the  F435W and F606W filters.  The circle shows the relative
HST/Chandra error circle with a radius of $0.3\arcsec$.  There is one
star located well within the error circle (star A) and one located near
the edge (star B).  Star A has a position of $\alpha=$ 02h 43m 38s.163
and $\delta=$ +01$\arcdeg$ 24$\arcmin$ 11$\arcsec$.173 (J2000).  Star B
has a position of $\alpha=$ 02h 43m 38s.145 and $\delta=$ +01$\arcdeg$
24$\arcmin$ 11$\arcsec$.438 (J2000).  Neither source appears spatially
extended.  The ST magnitudes of Star A uncorrected for reddening are
$25.66 \pm 0.05$ in the F435W filter and $26.50 \pm 0.06$ in the F606W
filter.  For Star B, the ST magnitudes are $26.71 \pm 0.09$ in the
F435W filter and $28.00 \pm 0.19$ in the F606W filter.  

To compare the colors of the counterparts to the X-ray source with
theoretical predictions, some of which are not available in HST/ACS or
HST/WFPC2 bands, we calculate the equivalent B and V magnitudes.  The
dereddened V and B magnitudes of the two stars, calculated using the
reddening correction described above and using the spectrum of an A0V
star, are listed in Table~\ref{optcounter}.  There may be additional
reddening within NGC 1073 or internal to the source which would affect
these values.  Because the B-band is not a precise match to the F435W
and the F606W band is significantly broader than the standard V band,
the shape of the spectrum used to calculate the ACS/WFC will affect the
magnitude translation.  To investigate the magnitude of this effect, we
re-calculated the magnitudes using spectra for O5V, B5V, and A5V stars
in addition to the A0V star.  These stars cover a color range in
$(B-V)_{0}$ from --0.33 to +0.15.  We find that the band translation
contributes an error of $\pm 0.05$ in B and $\pm 0.04$ in V.  These
band translation uncertainties have been added linearly to the
photometry errors to find the total errors quoted on the magnitudes.

We also calculated dereddened ST magnitudes in the WFPC2 bands for the
two candidate counterparts, see Table~\ref{optcounter}.  The
uncertainty due to the band conversion was calculated as in the
previous paragraph.  There may be additional reddening within NGC 1073
or internal to the source which would affect these values.

\section{Discussion}

IXO 5 lies near a ring of star formation shown in Fig.~\ref{sb_image}. 
This ring has a diameter of about $20\arcsec$ or 1.6~kpc at the
distance to NGC 1073.  The H{\sc ii} region identified by \citet{arp04}
as a possible counterpart lies $8\arcsec$ to the SE of IXO 5 and can be
seen in the lower left corner of Fig.~\ref{sb_image}.  This  H{\sc ii}
region is excluded as a counterpart to IXO 5, but may be part of the
same star-forming event that lead to the formation of the ring.  The
H{\sc ii} region has a redshift consistent with that of NGC 1073.  

A color-magnitude diagram for the objects in the ring of star formation
is shown in Fig.~\ref{sb_hr}.  As noted above, some of the objects on
the diagram may be unresolved or marginally resolved H{\sc ii} regions
or star clusters.  For comparison, we have plotted the isochrones for
8~Myr and 16~Myr old stellar populations with solar metallicity, $z =
0.019$ and $Y=0.273$, from \citet{bertelli94} and updated and cast into
the HST/WFPC2 bands by \citet{girardi02}.  The data were obtained from
the Padova web site (http://pleiadi.pd.astro.it/).  The overall
distribution of objects on the diagram is consistent with that expected
for a young stellar population.  The ring of star formation is similar
to that found near the source X7 in NGC 4559 \citep{soria04}.  This
strengthens the idea that the bright X-ray sources found in normal
spiral galaxies are associated with recent star formation events.

\begin{deluxetable}{ccc}
\tablecaption{Optical stars near IXO 5 \label{optcounter}}
\tablewidth{0pt}
\tablehead{
\colhead{Magnitude} & \colhead{Star A} & \colhead{Star B}}
\startdata
$m_{\rm F435W}$     & $25.66 \pm 0.05$ & $26.71 \pm 0.09$ \\
$m_{\rm F606W}$     & $26.50 \pm 0.06$ & $28.00 \pm 0.19$ \\ \hline
$(m_{\rm F439W})_0$ & $25.44 \pm 0.10$ & $26.49 \pm 0.14$ \\
$(m_{\rm F606W})_0$ & $26.42 \pm 0.08$ & $27.92 \pm 0.21$ \\
$(m_{\rm F439W}-m_{\rm F606W})_0$
                    & $-0.98 \pm 0.18$ & $-1.43 \pm 0.34$ \\
$V$                 & $26.23 \pm 0.10$ & $27.73 \pm 0.23$ \\
$B-V$               & $ 0.04 \pm 0.14$ & $-0.41 \pm 0.27$ \\
$V_0$               & $26.11 \pm 0.10$ & $27.61 \pm 0.23$ \\
$M_{V}$             & $-4.96 \pm 0.41$ & $-3.46 \pm 0.46$ \\
$(B-V)_0$           & $ 0.00 \pm 0.14$ & $-0.45 \pm 0.27$ \\
$\xi$               & $20.3 \pm 0.2$   & $21.3 \pm 0.2$ \\
\enddata

\vspace{-12pt}\tablecomments{Candidate optical counterparts to
ultraluminous X-ray source IXO 5.  The first two rows give the measured
quantities which are the ST magnitudes with no correction for reddening
in the WFC/ACS F435W and F606W bands.  The other rows give transformed
magnitudes derived from the measured values.  These are:  the
dereddened ST magnitudes in the WFPC2 F439W and F606W bands, the colors
derived from these ST magnitudes, the absolute V magnitude corrected
for reddening and assuming a distance of 16.4~Mpc the reddening 
corrected B--V color, and the X-ray to optical flux ratio, defined as
$\xi = B_{0} + 2.5 \log F_{X}$ where $F_{X}$ is the X-ray flux density
at 2~keV in $\mu$Jy \citep{vp95}.} \end{deluxetable}

At a distance of 16.4~Mpc, the luminosity of IXO 5 (assuming isotropic
emission) is $2.2 \times 10^{39} \rm \, erg \, s^{-1}$ in the
0.3--7~keV band.  This luminosity is not above the peak luminosities of
stellar-mass black hole candidate X-ray transients found in the Milky
Way and within the Eddington limit for a `normal', i.e.\ less than $20
M_{\odot}$, black hole.  Therefore, the system is likely an X-ray
binary containing a `normal' black hole rather than intermediate mass
black hole.  However, this luminosity is consistent with that found
from the ROSAT observation made during 14-17 Jan 1998
\citep{colbert02}, which is six years before the Chandra observation.
The constancy suggests that the source is persistently bright, unlike
the Milky Way black hole candidate X-ray transients.  The source is
significantly brighter than any persistent black hole candidate X-ray
binary in the Milky Way or the Magellanic Clouds, the most luminous of
which is LMC X-1 at $3 \times 10^{38} \rm \, erg \, s^{-1}$
\citep{gierlinski01}.  Therefore, it appears to represent a different
sort of X-ray binary, although perhaps only in evolutionary state, from
those found nearby.

We found two candidate optical counterparts to IXO 5 with optical
magnitudes presented in  Table~\ref{optcounter}.  We calculated the
absolute magnitudes of the stars assuming a distance of 16.4~Mpc with
an uncertainty of 20\%.  If the optical light comes only from the
companion star, then the absolute magnitude and $(B-V)_{0}$ color of
the candidate counterparts can be used to classify them.  Using the
tables by Schmidt-Kaler in \citet{aller82}, we find that the possible
spectral types for star A are all early supergiants, B5Ib to A8Ib.  
For star B, the possible spectral types are B0V-B1V, B1IV-B2IV,
B2III-B3III, and B5II. This spectral classification assumes that the
optical light is dominated by the companion star with little or no
contribution from the accretion disk.  

To help understand if reprocessed emission should be important, we
calculate the X-ray to optical flux ratio, defined following
\citet{vp95} as $\xi = B_{0} + 2.5 \log F_{X}$ where $F_{X}$ is the
X-ray flux density at 2~keV in $\mu$Jy.  For star A, $\xi = 20.3$ which
is higher than the X-ray to optical flux ratio for any high-mass X-ray
binary within the Milky Way or Magellanic clouds except for LMC X-3. 
The value is consistent with that found for low-mass X-ray binaries,
but somewhat lower than average.  The $(B-V)_{0}$ color of $0.0 \pm
0.14$ is also consistent with those of low-mass X-ray binaries,
indicating that light from an accretion disk could contribute to the
optical emission.  For star B, the X-ray to optical flux ratio is
21.3.  This is higher than the ratios of high-mass X-ray binaries and
consistent with the ratios of low-mass X-ray binaries where the
accretion disk dominates over the light from the companion star.  The
color, $(B-V)_{0} = -0.45 \pm 0.21$, of star B is at the blue end of
the color distribution of low-mass X-ray binaries \citep{vp95}.  We
conclude that reprocessed emission likely contributes to the optical
light of either candidate.  This casts doubt on the spectral
classifications derived above and suggests that the effect of the X-ray
emission should be taken into account to understand the optical light.

\citet{rappaport05} have calculated  binary evolution models for black
hole X-ray binaries and made predictions concerning the optical
properties including contributions from both the companion star and the
accretion disk.  While the applicability of such models to
ultraluminous X-ray sources with higher luminosities ($L_X \sim 10^{40}
\rm \, erg \, s^{-1}$) must remain in question because of concerns with
exceeding the Eddington limit for `normal' stellar-mass black holes,
the models are clearly applicable to this system.  Comparing the
properties of the two candidate optical counterparts with those
predicted for black hole X-ray binaries, i.e.\ Figure~12 in
\citet{rappaport05}, we find that star B is bluer than any of the model
binaries consistent with its $M_V$.  Star A is consistent with the
$M_V$ and $B-V$ color predicted for binaries with donor stars with
original masses of $6-8 M_{\odot}$.  However, the binary evolution
models appear to underestimate the X-ray luminosity.  If the age of the
X-ray binary is similar to the 8--16~Myr age of the star forming ring,
then the predicted luminosities for binaries with original donor masses
of  $6-8 M_{\odot}$ are all well below $10^{39} \rm \, erg \, s^{-1}$.
This may indicate some deficiency in the modeling.  Binaries in this
mass range may undergo mass transfer at super-Eddington rates as the
donor ascends the giant branch, but this would not be expected to occur
until $\sim 100$~Myr after the formation of the binary.  Also, in this
case the X-ray luminosity would need to be restricted by the Eddington
limit to reproduce the observed luminosity.

\section*{Acknowledgments}

PK thanks Steve Spangler and Sarah Iverson for useful discussions, the
Aspen Center for Physics for its hospitality during the workshop
``Compact Objects in External Galaxies'', and the referee for comments
which improved the paper.  PK acknowledges partial support from STScI
grant HST-GO-10001.02-A and Chandra grant CXC GO4-5086X.  STSDAS and
PyRAF are products of the Space Telescope Science Institute, which is
operated by AURA for NASA.


\label{lastpage}

\end{document}